\documentclass[preprint]{revtex4-1}

\usepackage[utf8]{inputenc}
\usepackage{amsmath}
\usepackage{amsfonts}
\usepackage{amssymb}
\usepackage{graphicx}
\usepackage{dcolumn}
\usepackage[pdftex,bookmarks=true,backref=page,pagebackref=false,colorlinks=true, linkcolor=blue]{hyperref}

\usepackage{changes}
\usepackage{color}
\usepackage{outlines} 

\newcommand{\orcid}[1]{\href{https://orcid.org/#1}{\,\protect\includegraphics[width=8pt]{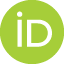}}}

\newcommand{\code}{M3D-$C^1$}

\begin{document}

\title{ITER cold VDEs in the limit of a perfectly conducting first wall}

\author{C F Clauser \orcid{0000-0002-2597-5061}} 
\email{cclauser@pppl.gov}

\author{S C Jardin \orcid{0000-0001-6390-6908}}
\affiliation{Princeton Plasma Physics Laboratory, Princeton, NJ 08543, United States of America}

\date{\today}

\begin{abstract}
Recently, it has been shown that a vertical displacement event (VDE) can occur in ITER even when the walls are perfect conductors, as a consequence of the current quench [A. H. Boozer, \href{https://aip.scitation.org/doi/10.1063/1.5126339}{Physics of Plasmas {\bf 26} 114501 (2019)}].
We used the extended-MHD code \code with an ITER-like equilibrium and induced a current quench to explore cold VDEs in the limit of perfectly conducting walls, using different wall geometries. In the particular case of a rectangular wall with the side walls far away from the plasma, we obtained very good agreement with the analytical model developed by Boozer that considers a top/bottom flat-plates wall.
We show that the solution in which the plasma stays at the initial equilibrium position is improved when bringing the side walls closer to the plasma.
When using the ITER first wall in the limit of a perfect conductor, the plasma stays stable at the initial equilibrium position far beyond the value predicted by the flat-plates wall limit.
On the other hand, when considering the limit in which the inner shell of the ITER vacuum vessel is acting as a perfect conductor, the plasma is displaced during the current quench but the edge safety factor stays above $2$ longer in the current decay compared to the flat-plates wall limit.
In all the simulated cases, the vertical displacement is found to be strongly dependent on the plasma current, in agreement with a similar finding in the flat-plates wall limit, showing an important difference with usual VDEs in which the current quench is not a necessary condition.
\end{abstract}


\maketitle

\section{Introduction}\label{Sec:Intro}

Vertical Displacement Events (VDEs) are major disruptions that occur in elongated tokamak plasmas. Reliable simulations to predict possible scenarios in ITER are essential. In a ``cold VDE'', the disruption thermal quench occurs when the plasma is still at the equilibrium position, and then the plasma displaces upward or downward (see for example, Ref. \onlinecite{Boozer2012}).  This is in contrast to a ``hot VDE'' in which the hot plasma column displaces vertically due to loss of position control, and the thermal quench occurs later as a result of the plasma contacting the first wall.

In ITER, the first wall is expected to have poor toroidal conductivity since it is made of panels with gaps between them \cite{Raffray2014}. However, it has been pointed out\cite{Roccella2016a} that if the plasma fills those gaps during a disruption event, it might short circuit the first wall making it toroidally conducting.  
Recently, Boozer \cite{Boozer2019} has considered the limiting case in which the ITER first wall acts as a perfect conductor and, approximating it as two top/bottom flat plates, has shown that, even in this perfectly conducting wall limit, a current quench could drive a VDE in which active controls would not be applicable since the magnetic fields cannot to penetrate the wall in the required time-scale. In that simple model, which will be referred here as the \textit{flat-plates wall limit}, the edge safety factor could drop down to values around $q_{95}\sim 2$ even when the plasma current is still large. This condition could lead to a large and unacceptable halo currents and local force densities in the first wall \cite{Boozer2019}.

Here, we use the \code code to simulate cold VDEs in the limit of perfectly conducting walls. We consider different geometries and show that in this limit, the ITER first wall keeps the plasma centered far beyond the value predicted by the flat-plates wall limit. However, when assuming that the perfectly conducting structure is the inner shell of the ITER vacuum vessel, the plasma is displaced during the current quench but the safety factor should not decrease as fast as in the flat-plates wall limit.
 
This paper is organized as follows: In Section \ref{Sec:Model} we present the plasma equilibrium and the different wall geometries employed. In Sec. \ref{Sec:flat_plates_limit} we show the limiting case in which the flat-plates wall limit is applicable. Finally, we show in Sec. \ref{Sec:ITERcase} the results using the ITER first wall as a perfect conductor as well as the case in which the inner shell of ITER vacuum vessel acts as a perfect conductor. 

\section{Numerical Model}\label{Sec:Model}

We started the simulations from a standard ITER-like equilibrium with plasma current $I = 15\,$MA, vacuum magnetic field $B_0 = 5.3\,$T, internal inductance $l_i(3) = 0.816$, magnetic axis $(R_m, Z_m) = (6.524\,\text{m}, 0.537\,\text{m})$, an active lower x-point at $(R, Z) = (5.148\,\text{m}, -3.386\,\text{m})$ and a passive upper x-point at $(R, Z) = (6.618\,\text{m}, 4.748\,\text{m})$. The equilibrium is the same as the one used in Ref. \onlinecite{Clauser2019}.
\begin{figure*}
\includegraphics[width=0.9\columnwidth]{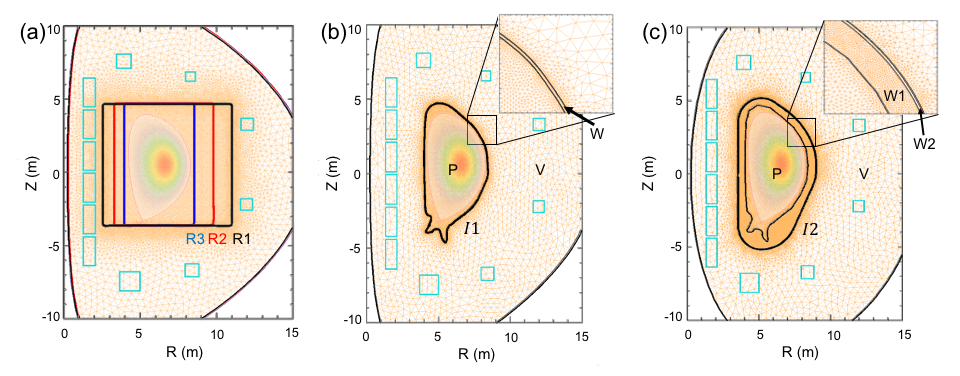}
\caption{\label{Fig:ITERmesh} Different wall geometries, computational domain and meshes employed for the simulations. In all cases, the computational domain includes a plasma region (P), a wall region (W) and a vacuum region (V). (a) shows three different rectangular walls labeled as R1, R2 and R3, (b) shows a thin ITER first wall model $I1$ and (c) shows a thick ITER vessel model, $I2$, with two wall layers (W1 and W2).}
\end{figure*}

\code \cite{Breslau2009,Jardin2012} is a high order finite-element code that solves the extended-MHD equations in a plasma region (P). It has also been extended to include a (resistive) wall region (W) and a vacuum region (V) \cite{Ferraro2016}, and external coils are included to calculate the initial equilibrium. These regions and the meshes are shown in  Fig. \ref{Fig:ITERmesh}. The code was successfully employed to study VDEs in tokamaks in both 2D and 3D \cite{Pfefferle2018,Clauser2019}. In support of this capability, a benchmark in 2D with well-known codes NIMROD \cite{Sovinec2004} and JOREK \cite{Holzl2012} was recently carried out with success \cite{Krebs2020}. 
In this study, we use the 2D version of \code and we employed different wall geometries: (a) a thin rectangular wall, (b) a thin ITER first wall and (c) a thick two-layer ITER wall. We note here that \code can handle only closed contours around the plasma as a resistive wall. Note also that the ``thin'' walls were of thickness $2-6\,$cm with finite elements in the wall.

For the rectangular geometry, Fig. \ref{Fig:ITERmesh}(a), we scanned over three different cases  R1, R2 and R3. The former with the side walls far away from the plasma but inside the coils region, and the latter with the side walls very close to the plasma. The wall resistivity employed in the rectangular geometries as well as in the thin ITER wall geometry, Fig. \ref{Fig:ITERmesh}(b), was $\eta_w = 3.6\times10^{-8}\,\Omega\text{m}$ which is similar to the resistivity of Beryllium and approximates an ideal conducting wall on the timescales of these calculations. 

The last case, Fig. \ref{Fig:ITERmesh}(c), shows a new capability that has been recently incorporated in \code: a resistive wall region with different layers and anisotropic wall resistivities. This new model is an important step towards more realistic simulations of resistive wall instabilities. As an approximate model of the first wall panels and blanket modules, the W1 region has a poor toroidal conductivity ($\eta_{w1}^{(tor)} = 1.933 \times 10^{-2}\,\Omega\text{m}$) but very good poloidal conductivity ($\eta_{w1}^{(pol)} = 1 \times 10^{-8}\,\Omega\text{m}$).  The W2 region constitutes the ITER inner vacuum vessel shell, which is a very good conductor. Here we evaluate the limit when W2 approximates an ideal conductor,  setting an isotropic resistivity of $\eta_{w2} = 3.6\times10^{-8}\,\Omega\text{m}$.

All these simulations were performed in 2D using the single temperature MHD model described in Ref. \onlinecite{Ferraro2019}. To facilitate the discussion in the following sections, we introduce here the primary terms in the temperature equation that is solved in the plasma region:
\begin{subequations}
\begin{align}
\frac{3}{2} n \frac{\partial T}{\partial t} & = - \nabla \cdot {\bf q} + \eta {\bf J}^2 + (...) \label{Eq:dT_dt} \\
{\bf q}  & = - \kappa_\perp \nabla_\perp T - \kappa_\parallel \nabla_\parallel T .
\end{align}
\end{subequations}
Here, $\eta\sim T^{-3/2}$ is the Spitzer resistivity, which can be scaled by an arbitrary constant, and ${\bf q}$ is the heat flux. The ratio of parallel to perpendicular heat flux coefficients, $\kappa_\parallel/\kappa_\perp$, was set to $10^4 - 10^6$, large enough to minimize the magnitude of halo currents, which were not considered in Ref. \onlinecite{Boozer2019}.

\section{Top-bottom flat plates wall limit}
\label{Sec:flat_plates_limit}

Here we summarize some of the results presented in Ref. \onlinecite{Boozer2019} that will be useful in this work. Following that model, which we will refer as the flat-plates wall limit, the vertical displacement can be written as
\begin{subequations} 
\label{Eq:flat-plates}
\begin{align} 
& \delta = 0      & (I > I_*) \\
& \frac{\delta^2}{b^2} = 1.2158 \left( \frac{I_*}{I} - 1 \right)  & (I<I_*),
\label{Eq:delta}
\end{align}
\end{subequations}
where $\delta = Z_{m}(t) - Z_{m}(0)$ is the displacement of the magnetic axis measured from its initial equilibrium position and $b$ is the top/bottom wall distance measured from the same equilibrium magnetic axis position.
Equation \eqref{Eq:delta} has real solution when the plasma current has decayed below a critical current value $I_*$. When $I>I_*$ the solution is stable at $\delta = 0$. We can compute this critical current value by specifying the height $b_{x0}$ of the x-point, at which the radial magnetic field vanishes, obtaining
\begin{equation}
\frac{I_0}{I_*} = 1.2337 \frac{b_{x0}^2}{b^2}.
\end{equation}
This result shows that $I_* < I_0$ only when the ratio $b_{x0}/b > 0.9$. 

\begin{figure}
\includegraphics[width=1.0\columnwidth]{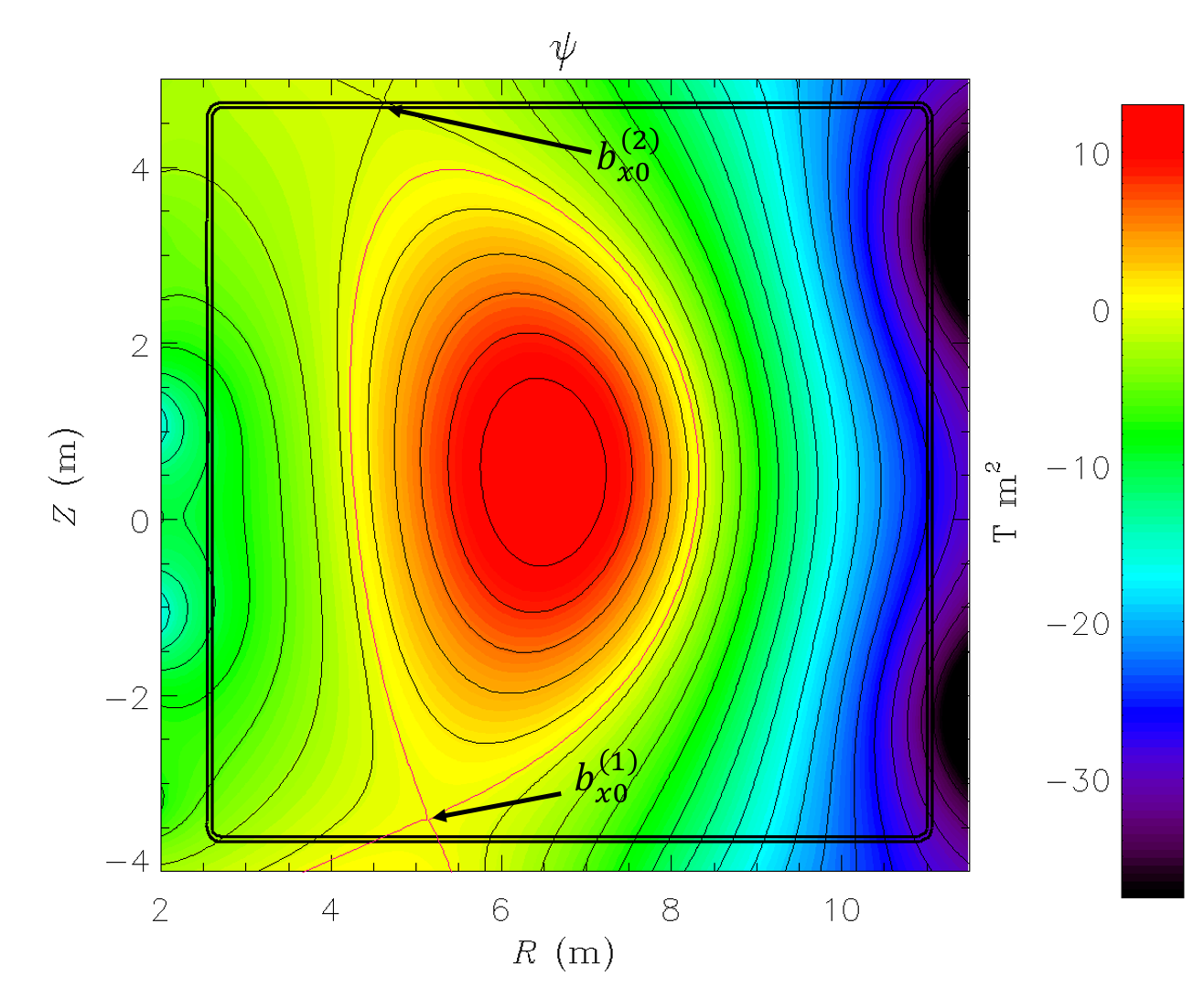}
\caption{\label{Fig:Equil_rect} ITER equilibrium poloidal flux employed is identical in all the cases, but different wall geometries were employed  (as indicated in Fig. \ref{Fig:ITERmesh}). Here we show with the R1 model which was used to approximate the two flat perfect conducting plates.}
\end{figure}

Therefore, in order to observe in our simulations a transition in which the plasma changes its equilibrium position from $\delta = 0$ to $|\delta| \neq 0$ when the plasma current decays below $I_*$, we have to choose the top/bottom walls close enough to the x-points. In addition, to make the comparison with the flat-plates wall limit, the top/bottom walls should be both equidistant to the equilibrium magnetic axis z-position $Z_{m}(0)$ and the side walls should be as far as possible from the plasma. That is why we used the R1 geometry shown in Fig. \ref{Fig:ITERmesh}(a). Figure \ref{Fig:Equil_rect} shows the equilibrium poloidal flux $\psi$ with contour lines. It also shows the two, active and passive, x-points. The theoretical model presented in Ref. \onlinecite{Boozer2019} assumes up/down symmetry but the ITER equilibrium is not.
From Fig. \ref{Fig:Equil_rect} we get that $b_{x0}^{(1)}/b = 0.95$ and  $b_{x0}^{(2)}/b \approx 1$, leading to $I_{*}^{(1)} = 13.5\,$MA and $I_{*}^{(2)} = 12.2\,$MA, respectively. From the analysis performed for the symmetric case, we infer from this that the configuration will certainly be unstable when the plasma current decays below $12.2\,$MA. 

\begin{figure}
\includegraphics[width=0.9\columnwidth]{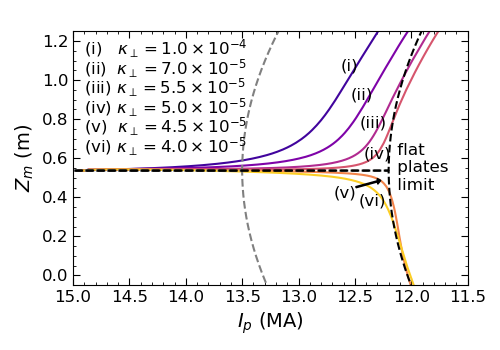}
\caption{\label{Fig:100} $Z-$coordinate of the magnetic axis as a function of the plasma current for different perpendicular heat flux coefficients $\kappa_\perp$. The dash curves show the flat-plates wall limit model, Eqs. \eqref{Eq:flat-plates}.}
\end{figure}

\begin{figure}
\includegraphics[width=0.9\columnwidth]{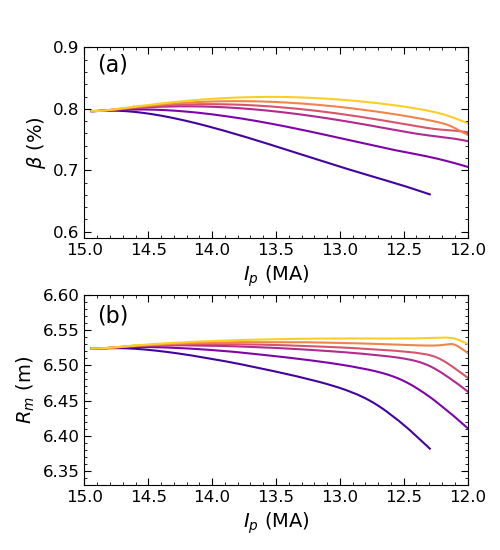}
\caption{\label{Fig:101} (a) Plasma $\beta$ and (b) $R-$coordinate of the magnetic axis as a function of the plasma current for different perpendicular heat flux coefficient $\kappa_\perp$, as indicated in Fig. \ref{Fig:100}.}
\end{figure}

In order to compare with the flat-plates wall limit model, we evolved this initial equilibrium increasing the plasma Spitzer resistivity by $10^5$ to simulate the current quench. Figure \ref{Fig:100} shows the magnetic axis Z-position as a function of the plasma current, for slightly different but small perpendicular heat flux coefficients $\kappa_\perp$ (to get $\kappa_\perp$ in SI units, multiply by $1.542\times 10^{26} \,\text{m}^{-1}\text{s}^{-1}$).
The artificial increase in the Spitzer resistivity produces a current quench (CQ) without a thermal quench (TQ). This keeps the plasma $\beta$ approximately constant, avoiding the inward displacement and initial kick that triggers the VDE (as used, for example, in Ref. \onlinecite{Clauser2019}). The plasma $\beta$ and $R-$coordinate of the magnetic axis, $R_m$ are shown in Fig. \ref{Fig:101} as a function of the plasma current. The perpendicular heat flux coefficient $\kappa_\perp$ was slightly varied in order to scan over small $\beta$ variations, primarily due to the increased ohmic heating when the resistivity is increased.
We can observe from Fig. \ref{Fig:100} that the simulation is in excellent agreement with the flat-plates wall limit model. It was not possible to generate a case that surpasses this stability limit using the R1 rectangular wall.

\begin{figure}
\includegraphics[width=0.9\columnwidth]{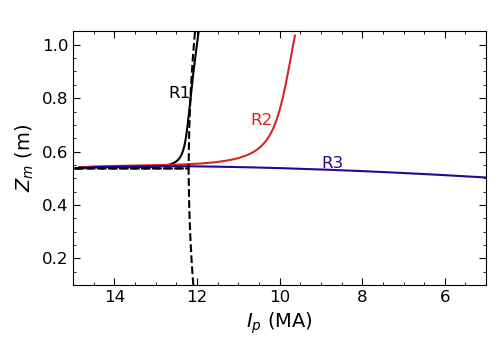}
\caption{\label{Fig:105} $Z-$coordinate of the magnetic axis as a function of the plasma current for different rectangular walls, as shown in Fig. \ref{Fig:ITERmesh}(a).}
\end{figure}

However,  if the side walls are brought closer to the plasma the stability of the $\delta=0$ solution improves drastically. Figure \ref{Fig:105} shows the case ``(iv)'' of Fig. \ref{Fig:100} ($\kappa_\perp = 5\times 10^{-5}$) for the R1 model but now compared with the R2 and R3 models.
The side walls, which were neglected in Ref. \onlinecite{Boozer2019}, are seen to be very stabilizing to the VDE. They impose a constraint on the normal component of the perturbed magnetic field and hence on the plasma motion.

\section{Cold VDE driven by a Thermal Quench}

\begin{figure}
\includegraphics[width=0.9\columnwidth]{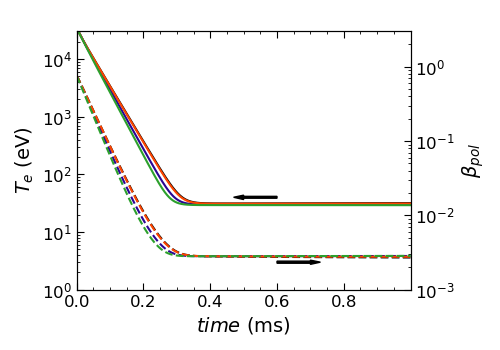}
\caption{\label{Fig:200} Central electron temperature and plasma poloidal beta as a function of time for different wall geometries.}
\end{figure}

In a more realistic situation, a  cold VDE is triggered by a thermal quench (TQ). This produces a collapse in the plasma $\beta$ and temperature, as shown in Fig. \ref{Fig:200}, which usually triggers the vertical displacement.
Here, we have chosen a post-TQ $T_e\approx 30\,$eV as a standard post-TQ plasma core temperature. During this collapse, which in these simulations lasts $\lesssim 0.3\,$ms, the plasma changes its equilibrium state to satisfy the force-free condition and the evolution of this new state will be affected by the boundary condition imposed by the conducting walls. 
\begin{figure}
\includegraphics[width=0.6\columnwidth]{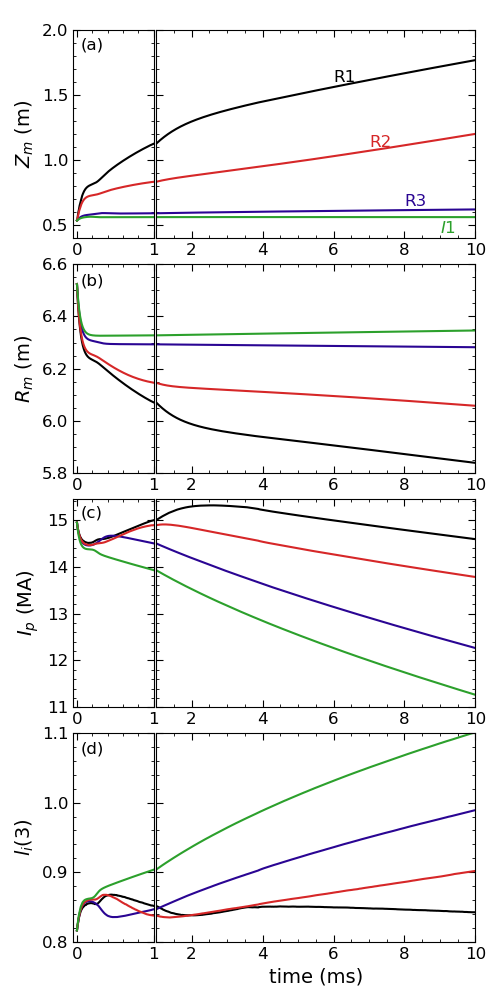}
\caption{\label{Fig:201} Global quantities as a function of time for different wall models: (a) $Z-$coordinate and (b) $R-$coordinate of the magnetic axis, (c) plasma current and (d) internal inductance. A zoom in time at the beginning shows the initial displacement caused by the thermal quench.}
\end{figure}
The evolution of some relevant global quantities is shown in Fig. \ref{Fig:201}. We can observe that in these cases, the R1 and R2 wall geometries lead to a significant upward motion, as shown in Fig. \ref{Fig:201}(a), while bringing the wall closer such as in R3 and $I1$ geometries substantially reduces the displacement. 

However, the nature of this upward/downward movement is different from that of a standard VDE, which exhibits exponential growth on the resistive wall timescale. Here, the vertical movement is a function of the plasma current and radial position. If the side walls are far enough away, such as in the R1 and R2 cases, the plasma will displace into a region where the external fields are more destabilizing. The combination of the stronger destabilizing force from the external fields and the weaker passive stabilization due to the reduced current will cause the configuration to seek a new equilibrium state, displaced vertically.  It is seen that the R1 and R2 configurations displace vertically even before the plasma current begins to decrease. Freezing the plasma current (at a given major radius position) at any later time leads to a constant $Z_m$ time evolution (an example is shown in the next section). In this sense, these simulations show an extension of the flat-plates wall limit for different wall geometries.

Figure \ref{Fig:201}(c) shows the current quench for all different cases. The current quench is not only affected by the plasma temperature after the TQ, which is almost identical in all cases ($\sim 30\,$eV) but also by the wall geometry. Different wall geometries allow the plasma to change its internal inductance in different ways, as shown in Fig. \ref{Fig:201}(d), in order to keep the magnetic fluxes constant and to satisfy their respective boundary condition.

\section{The ITER Case}
\label{Sec:ITERcase}
\begin{figure}
\includegraphics[width=0.9\columnwidth]{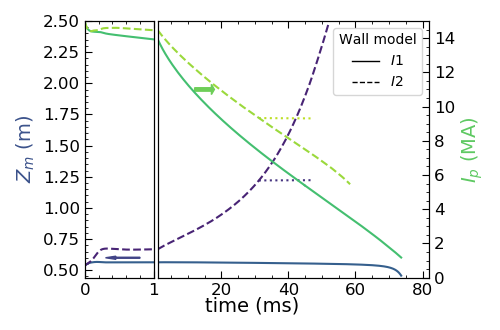}
\caption{\label{Fig:402} Time evolution of the $Z-$coordinate of the magnetic axis and plasma current for models $I1$ (solid) and $I2$ (dash). See Fig. \ref{Fig:ITERmesh} for models' reference. The dotted line shows that when freezing the current quench (by reducing the plasma resistivity), $Z_m$ also freezes in time.}
\end{figure}

\begin{figure}
\includegraphics[width=0.9\columnwidth]{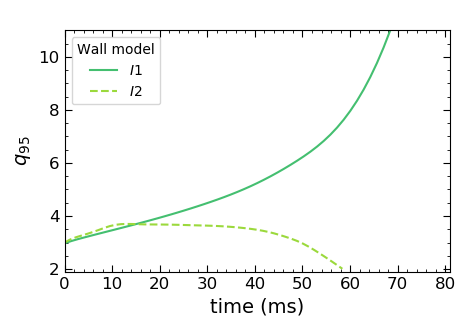}
\caption{\label{Fig:403} Edge safety factor ($q_{95}$) as a function of time for wall models $I1$ (solid) and $I2$ (dash). See Fig. \ref{Fig:ITERmesh} for models' reference.}
\end{figure}

Fig. \ref{Fig:201} shows that the perfectly conducting ITER first wall limit (model $I1$) is the most stable case we scanned. A longer time history of the $Z-$coordinate of the magnetic axis and plasma current is shown in Fig. \ref{Fig:402}. We can observe that the plasma is stable at $\delta = 0$ far beyond the flat-plates wall limit, and goes to a solution $|\delta|\neq 0$ when the plasma current is below $\sim 2\,$MA. This result leads to an edge safety factor increasing in time as shown in Fig. \ref{Fig:403}, improving the stability of non-axisymmetric modes.
On the other hand, we also include in Fig. \ref{Fig:402} the results from the thicker wall model $I2$ shown in Fig. \ref{Fig:ITERmesh}(c), in which the perfect conductor is assumed to be the outer wall layer $W2$ which accounts for the inner shell of the ITER vacuum vessel. Since this perfectly conducting wall is further away than in the model $I1$, we can observe that the plasma is displaced from the very beginning, as the plasma current decreases.
However, the vertical displacement is not as fast as in the flat-plates wall limit so that the edge safety factor, shown in Fig. \ref{Fig:403}, remains above $2$ until the current has decayed below $6\,$MA. The dotted line in Fig. \ref{Fig:402} exemplifies a case in which the current quench was frozen at an arbitrary time ($\sim 30\,$ms) by reducing the plasma resistivity. We can observe here that, as a consequence, $Z_m$ also freezes in time, illustrating that the $Z-$position is a function of the plasma current.

\section{Conclusions}
In this work, we used the extended-MHD code \code and explored the perfectly conducting first wall limit of an ITER cold VDE. We analyzed the flat-plates wall model by Boozer, performing a scan over different rectangular first wall geometries. We found that the model is applicable in a situation when the side walls are far enough from the plasma but bringing the walls closer considerably improves the stability of the $\delta = 0$ solution. It is shown that in the case of the ITER first wall the $\delta \neq 0$ solution takes place only when the plasma current has decayed below 2 MA. \\
Finally, when using the inner shell of the ITER vacuum vessel as a perfect conductor, the vertical displacement occurs at the time of the thermal quench but it is a slow enough function of the current decay, compared to the flat-plates wall limit, so that the edge safety factor does not decrease significantly down to values below $2$ until the current has decayed below $6\,$MA.\\
In all the simulated cases it was found that, in the limit of a perfectly conducting wall, the vertical displacement showed a strong dependency on the plasma current, in agreement with a similar finding in the flat-plates wall limit. Freezing the current leads to a freezing of the magnetic axis position in contrast to ``hot'' VDEs, which begin with full plasma current and beta values and do not occur in the limit of a perfectly conducting wall.

\begin{acknowledgments}
This work was supported by the U.S. Department of Energy under DOE Contract DE-AC02-09CH11466 and the SciDAC CTTS.
The authors are also grateful to Allen Boozer for fruitful discussions.
\end{acknowledgments}

\nocite{*}
\bibliography{bibliography}

\end{document}